\documentclass[letterpaper,10pt]{article} 

\usepackage{opticameet3} 

\newcommand\authormark[1]{\textsuperscript{#1}}

\usepackage{amsmath,amssymb}
\usepackage[colorlinks=true,bookmarks=false,citecolor=blue,urlcolor=blue]{hyperref} 
\usepackage{opticameet3} 
\usepackage{hyperref}
\usepackage{xspace}
\usepackage{amssymb}
\usepackage{amsmath,amssymb,amsfonts}
\usepackage{statmath}
\usepackage[figurename=Fig.]{caption}
\usepackage{algorithmic}

\hypersetup{
    colorlinks = false,
    hidelinks}

\usepackage{amsmath,amssymb}
\usepackage{subcaption}
\usepackage{siunitx}
\begin{document}

\title{Enabling Scalable Photonic Tensor Cores with Polarization-Domain Photonic Computing}


\vspace{-0.2 in}
\author{Amin Shafiee\authormark{1}, Linhong Chen\authormark{2}, Sudeep Pasricha\authormark{1}, Jie Yao\authormark{2*}, and Mahdi Nikdast\authormark{1*}}

\address{\authormark{1} Department of Electrical and Computer Engineering, Colorado State University, Fort Collins, CO, USA\\
\authormark{2}Department of Materials Science and Engineering, University of California, Berkeley, CA, USA\\}

\email{\authormark{1*}Mahdi.Nikdast@colostate.edu, \authormark{2*}yaojie@berkeley.edu} 

\copyrightyear{2025}
\vspace{-0.2 in}
\begin{abstract}
We present a silicon-photonic tensor core using 2D ferroelectric materials to enable wavelength- and polarization-domain computing. Results, based on experimentally characterized material properties, show up to 83\% improvement in computation accuracy compared to coherent networks. 
\end{abstract}

\vspace{-0.02in}
\section{Introduction}
\vspace{-0.08 in}

Photonic integrated circuits (PICs) have been deployed in a wide range of applications, from sensing and high-speed communication to energy-efficient optical computation in artificial intelligence (AI) accelerator systems. Conventional photonic AI accelerators rely on optical transmission levels and wavelength-domain computing, to encode data and perform matrix-vector or scalar multiplications in the optical domain. Such designs are often sensitive to accumulated optical noises and losses which directly affect the optical signal's amplitude, thereby lack the scalability required for performing more complex and computationally expensive tasks\cite{amin_jlt}. MX (M=Ge or Sn; X=Sn or S) materials are group IV monochalcogenide compounds that form a puckered layer structure similar to that of black phosphorus \cite{shi2018tin,jo2021photonic}. The remarkable characteristic of MX materials lies in their unconventional ferroelectric response with four ground states that allow ultrafast transitions between distinct crystalline structures with strong optical anisotropy. As a result, such transition enables in-plane rotation of optical axes, leading to distinctive optical properties. Prior work showed that the ferroelectric state of the MX materials can change by applying the electric field with magnitudes less than 0.5~V in less than 0.1~ns time due to reliance on the phonon's group velocity (which is in the order of the speed of sound) \cite{guan2022electric, shi2022role, ramesh2021electric}. In this paper, by careful experimental characterization of the optical properties of SnSe, we---for the first time to the best of our knowledge---design a photonic tensor core to perform matrix-vector multiplication relying on polarization of light rather than its amplitude using SnSe (a well-known MX material) ferroelectric properties integrated with silicon photonics platforms.\vspace{-0.1in} 

\begin{figure}[htbp]
    \centering
    \includegraphics[width=1\linewidth]{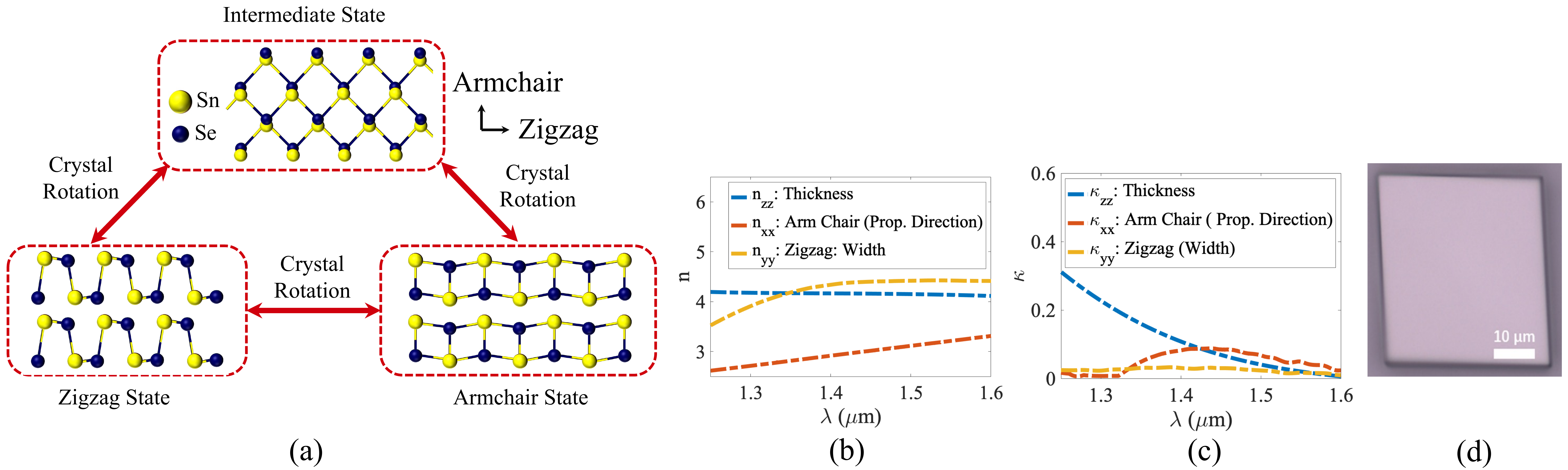}
    \vspace{-0.3in}
    \caption{(a) Molecular structure of SnSe in two nonvolatile ferroelectric states of armchair and zigzag. (b) Experimentally characterized (b) refractive index and (c) extiction coefficient of the SnSe in different in-plane crystalline orientations. (d) Microscopic image of the SnSe film from our experiments.}
    \label{fig1_intro_opt_prop}
     \vspace{-0.3in}
\end{figure}
   

\section{Design of Polarization-enabled Photonic Tensor Cores}
\vspace{-0.05 in}

The crystalline structure of two nonvolatile ferroelectric states along the armchair and zigzag directions of SnSe, the  MX material in this paper, is depicted in Fig.~\ref{fig1_intro_opt_prop}(a). 
 Assuming a linearly polarized light at the input, due to the birefringence property of SnSe in the different stable ferroelectric states when changing the crystalline orientation of SnSe, the Stokes parameters of the input light will change \cite{hanakata2016polarization}. 
The SnSe sample was fabricated using the physical vapor deposition (PVD) technique onto a clean Si/SiO$_2$ substrate. The optical transmission and reflection spectrum of the sample for zigzag and armchair directions was measured by changing the polarization of the input light. The measurement results were then combined with relative transmission fringe depth method (RTFD) and with our in-house solver based on the Levenberg-Merquardt method, which was used to calculate the complex refractive index profile of SnSe. The results for complex refractive index profile for SnSe is shown in Figs. \ref{fig1_intro_opt_prop}(b) and \ref{fig1_intro_opt_prop}(c). The microscopic image of the sample used for optical characterization is depicted in Fig. \ref{fig1_intro_opt_prop}(d).
\begin{figure}[htbp]
    \centering
    \includegraphics[width=1\textwidth]{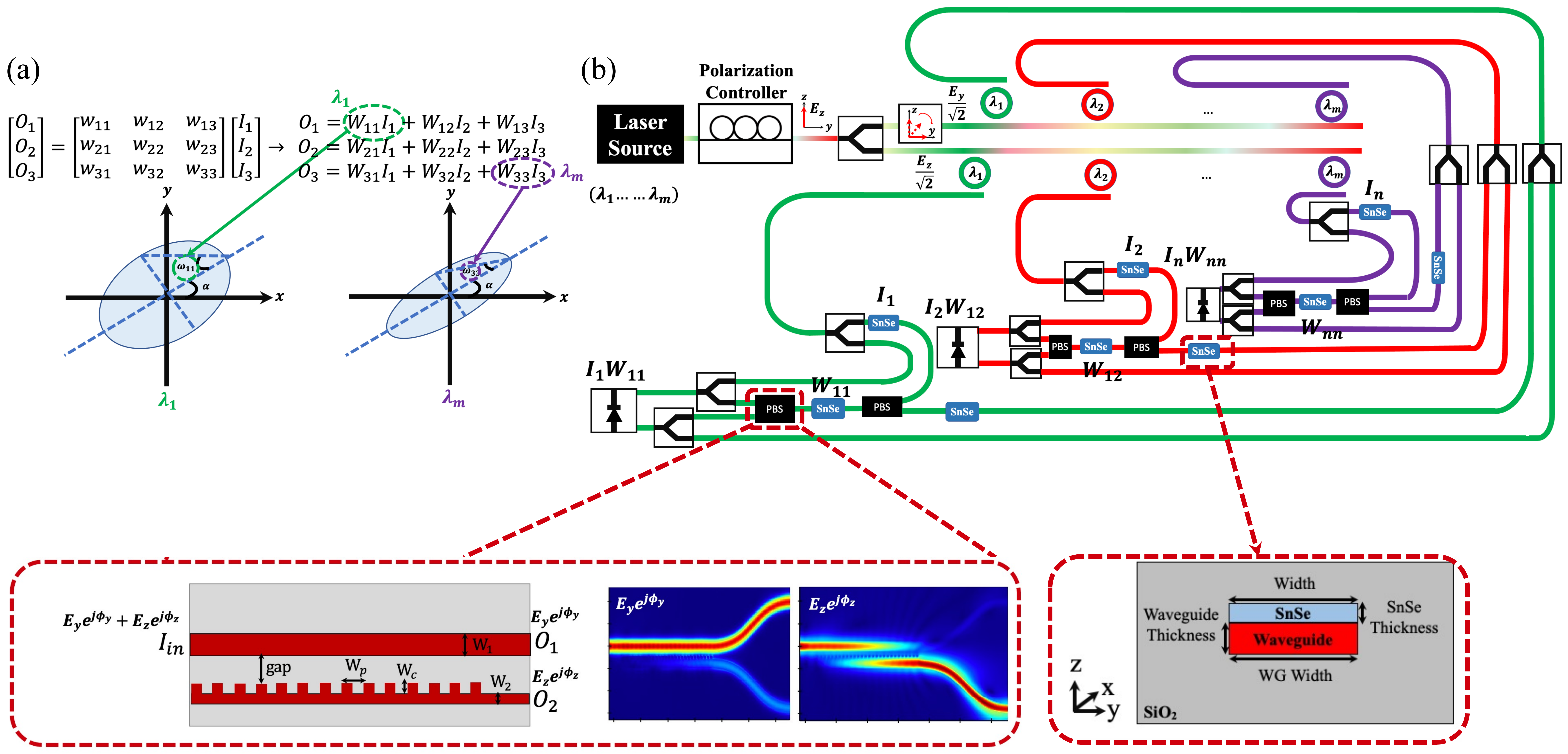}
    \vspace{-0.3 in}
    \caption{(a) Mapping the multiplication results onto Stokes parameters by changing the ellipticity angle of the polarization. (b) Design of the wavelength-multiplexed polarization-domain photonic tensor core.}
    \label{fig2_silicon}
     \vspace{-0.35in}
\end{figure}

We leverage the birefringence property of SnSe to realize, for the first time, photonic tensor cores with polarization-domain computing, capable of performing computationally expensive matrix-vector multiplications. The polarization ellipse is an instantaneous representation of a polarized electromagnetic wave (see Fig. \ref{fig2_silicon}(a)). The polarization ellipse can be characterized with Stokes parameters ($S_0$, $S_1$, $S_2$,$S_3$), where $S_0=E_y^2+E_z^2$, $S_1=E_y^2-E_z^2=cos(2\omega)cos(2\alpha)$, $S_2=2E_yE_zcos(\phi)=cos(2\omega)sin(2\alpha)$, and $S_3=2E_yE_zsin(\phi)=sin(2\omega)$. Here, $\omega$ is the ellipticity angle, $\alpha$ is the tilt angle, $E_y$ and $E_z$ are the magnitudes of the vertical and horizontal electric fields, and $\phi$ is the phase shift between the two vertical and horizontal components of the excited electric fields. Such a phase shift can be induced by careful manipulation of the crystal orientation of SnSe cell on top of the waveguide from zigzag, armchair, or any intermediate state \cite{hanakata2016polarization,ramesh2021electric}. From the Stokes parameters formulation, we can see that the scalar multiplication $A\times B=C$ can be performed by directly encoding $A$, $B$, and $C$ on the Stokes parameters of the polarized electromagnetic wave by changing $\phi$, which leads to a change in $\omega$.
The concept of the wavelength-multiplexed polarization-domain matrix-vector multiplication is depicted in Fig. \ref{fig2_silicon}(a). Observe that the element-wise multiplication results can be mapped on the Stokes parameters of the optical signals of different wavelengths assigned to the different multiplication results by manipulating its ellipticity angles. To realize a wavelength-multiplexed polarization-domain photonic tensor core (WPol-PTC) to carry out scalable matrix-vector multiplications using this scheme, we opted to design the architecture depicted in Fig. \ref{fig2_silicon}(b). The input optical signal has TM polarization with a broad wavelength range to be assigned to the multiplication results. Then, it will split by a 50:50 Y-branch. A polarization converter can be used on one of the outputs to excite TE0 ($E_y$) from TM0 ($E_z$ ). Next, the excited $E_y$ and $E_z$ of different wavelengths will be separated by sets of TE0 and TM0 micro ring resonators (MRRs). In this approach, the input vector elements ($I_1, I_2,...,I_n$) have been mapped on the Stokes parameters of the wave by two SnSe cells which act as phase shifters (PhS) to apply specific $\phi$ between the $E_y$ and $E_z$ components. Then, a broadband polarization beam combiner and splitter (PBS) to construct the polarized electromagnetic waves \cite{zhang2020ultra}. The ferroelectric state of the SnSe is used to tune the $\phi$ to map final multiplication result $W_{i,j}I_i$ onto the Stokes parameters of the propagating electromagnetic waves of different wavelengths\cite{hanakata2016polarization}. Note that the Stokes parameters of the electromagnetic waves are not directly measurable. Hence, we opted to use a PBS to separate $E_y$ and $E_z$ field components of polarized wave of a specific wavelength and TE- and TM-polarized y-branches, to detect their corresponding phase using optical interference.   



\vspace{-0.1in}

\section{Results and Discussions}
\vspace{-0.05in}

\begin{figure}[htbp]
    \centering
    \includegraphics[width=1\textwidth]{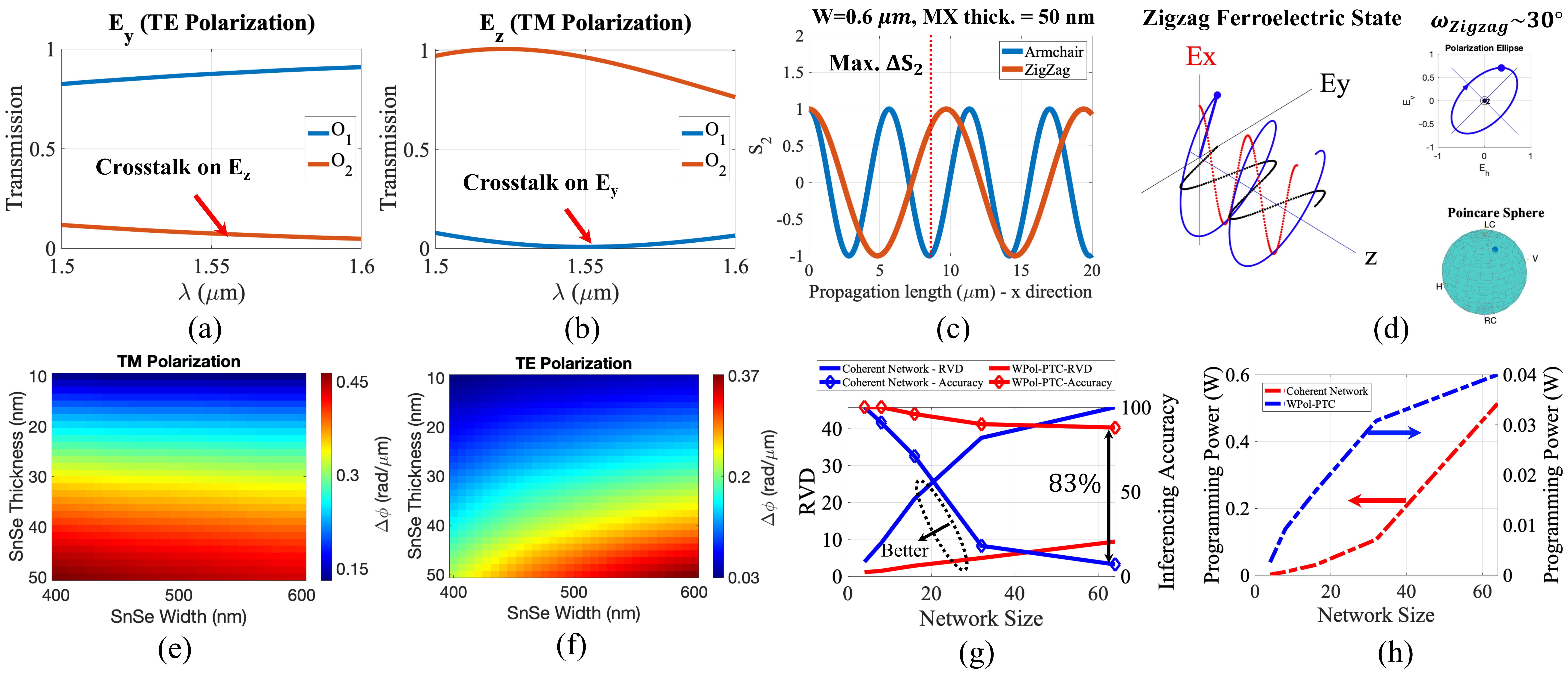}
    \vspace{-0.3 in}
    \caption{Optical transmission of PBS when (a) TE0 and (b) TM0 source is used. (c) $S_2$ of polarized light interacting with MX versus propagation length. (d) Polarization ellipse and electric field when  $\phi=$1~(rad). The design-space exploration of phase shift per unit length for (e) TM polarization and (f) TE polarization. (g) RVD and inferencing accuracy values for WPol-PTC and coherent networks of different sizes. (h) Programming power comparison between WPol-PTC and a MZI-based coherent network both with SnSe phase shifters.}
    \label{fig3_sin}
     \vspace{-0.35in}
\end{figure}
The optical transmission and crosstalk of the PBS is depicted in Figs. \ref{fig3_sin}(a) and \ref{fig3_sin}(b). In the PBS design (see the inset of Fig. \ref{fig2_silicon}(b)), the period of the grating region ($W_p$) is 280~nm with the total length of 6.72~$\mu$m with a duty cycle of 0.8, $W_1=$480~nm, $W_2=$425~nm, and $W_c=$130~nm. Observe that for fundamental TE0 source ($E_y$), up to 92\% of the electric field remains in the input waveguide and goes to output $O_1$. For the TM0 source ($E_z$), the field is coupled to the grating waveguide and goes to output $O_2$. The $S_2$ parameter of the polarized electromagnetic field when interacting with the SnSe on top of the waveguide as a function of the length of the cell (see the inset of Fig. \ref{fig2_silicon}) in both armchair and zigzag configuration is depicted in Fig. \ref{fig3_sin}(c). Observe that due to birefringence property of SnSe, the $S_2$ parameter will change at different rates depending on the ferroelectric state of SnSe. To maximize the $\Delta S_2$ between armchair and zigzag states (i.e., to maximize its range to map the multiplication results onto the stokes parameters), the length of the SnSe with the width of 600~nm and thickness of 50~nm is chosen to be 8.6~$\mu$m. Moreover, an example of polarization ellipse, its location on Bloch sphere, and the total electric field related to the electromagnetic wave where the phase shift between $E_y$ and $E_z$ is 1~rad (assuming SnSe is in the  zigzag state) is shown in Fig. \ref{fig3_sin}(d)ace explorat-on of SnSe-based PhS operating for TM ($E_z$) and TE ($E_y$) is depicted in Fig. \ref{fig3_sin}(e) and \ref{fig3_sin}(f), respectively. Observe that using SnSe to induce phase changes leads to a more compact, ultra-fast, and energy efficient PhS with minimum phase shifting length 8.5~$\mu$m for $E_y$ ($\approx$7~$\mu$m for $E_z$) and speed of less than 0.1~ns \cite{ramesh2021electric,shi2022role,guan2022electric,hanakata2016polarization}. 

To evaluate the performance of the designed WPol-PTC (see Fig. 2), we trained MZI-based coherent networks with Clements configuration of different sizes (4, 8, 16, 32, 64) on a linearly separable Guassian dataset, and captured the ideal weight matrices and non-ideal weight matrices in the presence of loss and crosstalk \cite{amin_jlt}. Then we captured the relative-variation distance (RVD) as well as the inferencing accuracy under the influence of loss and crosstalk according to the work in \cite{amin_jlt}. The same procedure was performed for WPol-PTC by considering the ideal and non-ideal weight matrix implemented by the network including the optical loss and crosstalk. Note that in our simulations, the optical insertion loss of PBSs is 0.24~dB and the worst-case crosstalk of 20\% for $E_z$ and $E_y$ electric field components on the PBS's outputs (see Fig. \ref{fig3_sin}(a) and \ref{fig3_sin}(b)) are included. The total loss of the SnSe cell is 0.1~dB/$\mu$m captured based on experimentally characterized optical properties of SnSe. The MRRs' worst-case drop port loss is considered to be 1.3~dB, experimentally characterized from the work in \cite{mirza2024experimental}. The RVD and inferencing accuracy results for the networks of different scales are reported in Fig. \ref{fig3_sin}(g). Compared to the coherent network, the accuracy values are significantly improved with maximum 10\% drop in their values when Stokes parameters were used to perform matrix-vector multiplication rather than the amplitude of the light. Lastly, considering SnSe-based PhS designed for TE modes in this paper to program the coherent network, and considering the resistance of the thin film SnSe for armchair and zigzag states ($\approx$880~$\Omega$ \cite{buruiana2022layered}) as well as maximum 0.5~V to change the its ferroelectric state from zigzag to armchair, the programming power to perform the matrix-vector multiplication on the trained weight matrices using the coherent network and WPol-PTC is reported in Fig. \ref{fig3_sin}(h). Observe that for the coherent network, the total programming power can be up to 0.8~W while for the WPol-PTC it is up to 40~mW when scaling up the network. Note that the static power consumption of SnSe cells is zero after programming (i.e., nonvolatile) \cite{ramesh2021electric}.
\vspace{-0.04in}
\vspace{0 in}

\vspace{-0.05 in}
\bibliographystyle{IEEEtran}
\bibliography{IEEEabrv,sample}

\end{document}